\documentclass[fleqn,usenatbib]{mnras}

\usepackage{mathptmx}
\usepackage[T1]{fontenc}
\usepackage{ae,aecompl}

\usepackage{graphicx} % Including figure files
\usepackage{amsmath} % Advanced maths commands
\newcommand{\angstrom}{\textup{\AA}}
\usepackage{amssymb} % Extra maths symbols
\usepackage{amsmath}
\usepackage[utf8]{inputenc}
\usepackage{mathptmx}

\title[Rapid Multiwavelength Timing of MAXI\,J1820+070]{A Black Hole X-ray Binary at $\sim$100 Hz: Multiwavelength Timing of MAXI\,J1820+070 with HiPERCAM and NICER}

\author[J. A. Paice et al.]{J. A. Paice$^{1,2}$\thanks{E-mail: j.a.paice@soton.ac.uk},
	P. Gandhi$^{1}$,
	T. Shahbaz$^{3,4}$,
	P. Uttley$^{5}$, 
	Z. Arzoumanian$^{6}$, \newauthor
	P. A. Charles$^{1}$\thanks{Leverhulme Emeritus Fellow},
	V. S. Dhillon$^{7,3}$,
	K. C. Gendreau$^{6}$,
	S. P. Littlefair$^{7}$,  
	J. Malzac$^{8}$,\newauthor
	S. Markoff$^{5}$,
	T. R. Marsh$^{9}$,
	R. Misra$^{2}$, 
	D. M. Russell$^{10}$,
	and A. Veledina$^{11,12}$.
	\\
	$^{1}$Department of Physics and Astronomy, University of Southampton, Highfield, Southampton, SO17 1BJ, UK\\
	$^{2}$Inter-University Centre for Astronomy and Astrophysics, Pune, Maharashtra 411007, India\\
	$^3$Instituto de Astrof\'\i{}sica de Canarias (IAC), E-38205 La Laguna,  Tenerife, Spain \\
    $^4$Departamento de  Astrof\'\i{}sica, Universidad de La Laguna (ULL),  E-38206 La Laguna, Tenerife, Spain \\
	$^{5}$Astronomical Institute 'Anton Pannekoek', University of Amsterdam, Science Park 904, NL-1098XH Amsterdam, the Netherlands\\
	$^{6}$X-ray Astrophysics Laboratory, Astrophysics Science Division, NASA’s Goddard Space Flight Center, Greenbelt, MD 20771, USA\\
	$^{7}$Department of Physics and Astronomy, University of Sheffield, Sheffield, S3 7RH, UK\\
	$^{8}$IRAP Universite de Toulouse, CNRS, UPS, CNES, Toulouse, France\\
	$^{9}$Astronomy and Astrophysics Group, Department of Physics, University of Warwick, Gibbet Hill Road, Coventry, CV4 7AL, UK\\
	$^{10}$New York University Abu Dhabi, PO Box 129188, Abu Dhabi, UAE\\
	$^{11}$Department of Physics and Astronomy, FI-20014 University of Turku, Finland\\
	$^{12}$Nordita, KTH Royal Institute of Technology and Stockholm University, Roslagstullsbacken 23, SE-10691 Stockholm, Sweden
}

\date{Submitted to MNRASL in original form 2019 Aug 08, Received 2019 Sept 19, Accepted 2019 Sept 20}

\pubyear{2019}

\begin{document}
	\label{firstpage}
	\pagerange{\pageref{firstpage}--\pageref{lastpage}}
	\maketitle
	
	\begin{abstract}
	{We report on simultaneous sub-second optical and X-ray timing observations of the low mass X-ray binary black hole candidate MAXI\,J1820+070. The bright 2018 outburst rise allowed simultaneous photometry in five optical bands ($ugriz_s$) with HiPERCAM/GTC (Optical) at frame rates over 100\,Hz, together with NICER/ISS observations (X-rays). Intense (factor of two) red flaring activity in the optical is seen over a broad range of timescales down to $\sim$\,10\,ms. Cross-correlating the bands reveals a prominent anti-correlation on timescales of $\sim$\,seconds, and a narrow sub-second correlation at a lag of $\approx$\,+165 ms (optical lagging X-rays). This lag increases with optical wavelength, and is approximately constant over Fourier frequencies of $\sim$\,0.3--10\,Hz. These features are consistent with an origin in the inner accretion flow and jet base within $\sim$\,5000 Gravitational radii. An additional $\sim$\,+5\,s lag feature may be ascribable to disc reprocessing. MAXI\,J1820+070 is the third black hole transient to display a clear $\sim$\,0.1\,s optical lag, which may be common feature in such objects. The sub-second lag {\em variation} with wavelength is novel, and may allow constraints on internal shock jet stratification models.}	
		
	\end{abstract}
	
	\begin{keywords}
		accretion, accretion discs -- X-rays: binaries -- X-rays: individual: MAXI J1820+070 -- stars: optical: variable -- black holes
	\end{keywords}
	
	%%%%%%%%%%%%%%%%%%%%%%%%%%%%%%%%%%%%%%%%%%%%%%%%%%
	
	%%%%%%%%%%%%%%%%% BODY OF PAPER %%%%%%%%%%%%%%%%%%

	\section{Introduction} \label{sec:Intro}
	
	%\citep{corral-santana_blackcat:_2016} is a citation (\citealt{shahbaz_evidence_2013}, while  \citet{sanchez_swift_2015} is another citation).
	
	%The day was 11th March 2018. An ordinary day, with an ordinary sun between ordinary clouds, while ordinary tea steamed from ordinary mugs. But it was not to last - nay, it was to change entirely. For you see, MAXI J1820+070 is an extraordinary source. Let me tell you...
	
	%Okay, actually put in a better intro paragraph at some point.

    Accreting black holes in binary systems are unrivalled laboratories for astrophysical conditions far beyond what can be reproduced on Earth. However, their small apparent angular sizes, prohibitively short time scales of flux variations, and unpredictable `outbursts' of enhanced accretion activity have historically made their study difficult. This is now beginning to change; the advent of a new generation of observatories, and using modes not possible before, allows us to probe them deeper than ever.
	
	ASASSN-18ey was first discovered on 2018 March 6 in the optical \citep{Tucker_ASASSN18ey_2018}, and then on March 11 classified as X-ray transient MAXI\,J1820+070 by \citet{ATel11399}. The source (hereafter \lq J1820\rq) quickly reached a flux of $\sim$4 Crab, making it one of the brightest X-ray transients ever (\citealt{corral-santana_blackcat:_2016}, \citealt{Shidatsu_Monitoring_2019}). Analysis of its optical/X-ray luminosity \citep{ATel11418}, X-ray power-law spectrum, measured disc blackbody temperature, and broadband timing power spectrum \citep{ATel11423} concluded that this source is a Low Mass X-ray Binary (LMXB). \citet{Torres_Dynamical_2019} dynamically confirmed a black hole (mass function $>$5.18 $\pm$0.15 M$_{\odot}$ and mass of $\sim$7.2 M$\odot$ with system inclination of 75$^{\circ}$), and its distance has been found to be $3.46^{+2.18}_{-1.03}$\,kpc \citep{Gandhi_GaiaDR2_2019}.
	%IF YOU HAVE SPACE, YOU CAN MENT$3.15^{+1.43}_{-1.57}$ION THAT THIS IS THE FIRST OBJECT WHOSE DISTANCE HAS BEEN ESTIMATED USING OPTICAL PARALLAX MEASUREMENTS.
	
	The origin of optical emission in LMXBs is generally considered to be a mixture of processes, including, e.g., X-ray reprocessing \citep{king_light_1998}, synchrotron radiation from a jet \citep{markoff_jet_2001, malzac_radiation_2018}, and/or an accretion flow (\citealt{fabian_gx339_1982}, \citealt{veledina_synchrotron_2011}). Fast timing observations can probe the interactions between these components and give important insight into the structure of the accretion flows. But such observations are challenging and only a handful of sources have been observed using {\em strictly simultaneous} rapid multiwavelength timing.
	
	To this end, we present simultaneous optical/X-ray timing results of J1820 from 2018 April 17 during its hard state \citep{Homan_J1820_2018}, carried out by the new HiPERCAM and NICER detectors at an unprecedented time resolution. %, and highlight interesting results that hint at the nature of the source.
	
	\section{Observations}

	\subsection{HiPERCAM/GTC -- Fast Optical Timing}
	\label{sec:HiPERCAM}
	
	%\begin{table*}
    %    \caption{Log of HiPERCAM observations.}
    %    \begin{minipage}{\textwidth}
    %    \centering
    %    \begin{tabular}{l c c c c c }
    %    \hline 
    %    UT date     & UT Start &   \# points & exp time (ms) &  cadence (ms) & median seeing (\arcsec)\\
    %    2018-04-17  & 03:25:55 &   3311528   & 1.98    &  2.93   & 2.2 \arcsec \\
    %    2018-04-19  & 05:11:50 &    457635   & 4.85    &  6.94   & XX \arcsec \\
    %    2018-05-20  & 05:06:41 &    397313   & 4.76    &  5.69   & XX \arcsec \\
    %    2018-05-22  & 05:11:49 &    164554   & 1.98    &  2.93   & XX \arcsec \\
    %    2018-06-07  & 04:33:08 &   1193768   & 1.98    &  2.93   & XX \arcsec \\
    %    \hline
    %    \end{tabular}
    %    \end{minipage}
    %    \label{table:rv}
    %\end{table*}

    %High-speed multi-colour photometry of J1820 was carried out during 5 epochs between 2018 April and June using HiPERCAM \citep{Dhillon_First_2018} on the 10.4 m Gran Telescopio Canarias (GTC).
    
    High-speed multi-colour photometry of J1820 was carried out using HiPERCAM \citep{Dhillon_First_2018} on the 10.4\,m Gran Telescopio Canarias. HiPERCAM uses 4 dichroic beamsplitters to image simultaneously 5 optical channels covering the $ugriz_{s}$-bands (respectively, wavelengths 3526, 4732, 6199, 7711 and 9156 \angstrom). The CCDs were binned by a factor of 8 and used in drift mode. We orientated the instrument (PA = 58$^{\circ}$) and used two windows (96x72 pixels each), one centered on J1820, and another on a comparison star, APASS--34569459 \citep{Henden_APASS_2015}. The observations discussed here were taken on 2018 April 17, from 03:26--06:11 UT, coordinated with NICER. The exposure time was 2\,ms, the cadence 2.9\,ms, the median seeing 2.\arcsec 2. The sky was affected by mild cirrus, but was reasonably photometric.
    
    %, ID 116612750537672500 in PanSTARRS DR1 \citep{Chambers_PanSTARRS_2016})
	
	We used the HiPERCAM pipeline software\footnote{\url{https://github.com/HiPERCAM/hipercam}} to de-bias, flat-field and extract the target count rates using aperture photometry with a seeing-dependent circular aperture tracking the centroid of the source. Sky background was removed using the clipped mean of an annular region around the target. The target was brighter than all stars in the field. We thus used the raw target counts for the analyses presented herein; note that our primary results are not affected when using photometry relative to the comparison star. 
%	Relative epldh, otoin the fimetry of J1820 was then carried out with respect to the comparison star to remove sky variations; however, the primary results presented herein are insensitive to this. %the comparison $u_{s}$ band was too faint to be useful, and thus the comparison $g_{s}$ band of the comparison was renormalised to the same level. %The average $ugriz_{s}$ magnitudes of J1820 were found to be 11.221, 11.192, 11.708, 11.844, 11.969 (with uncertainties of 0.069, 0.044, 0.033, 0.035, 0.042) respectively.
	
	%We did not divide by a comparison star due to the source being brighter than all stars in the field; instead, HiPERCAM/GTC's zero points were then used to calculate the magnitude, and atmospheric extinction was removed using \textsc{astroplan} \citep{Morris_Astroplan_2018} and parameters from \citet{King_Extinction_1985}. 

	\subsection{NICER -- X-ray}
	\label{sec:NICER}
	
	NICER (Neutron star Interior Composition ExploreR) is a new X-ray instrument aboard the International Space Station (ISS). It comprises 52 functioning X-ray concentrator optics and silicon drift detector pairs, arranged in seven groups of eight. Individual photons between 0.2-12 keV, and their energies, can be detected to a time resolution of 40 ns.
	
	%Due to flexible scheduling and good observing efficiency, NICER was able to observe the early period of J1820's outburst near-continuously for several weeks. Due to the low altitude of the ISS, the observations were frequently interrupted by occlusion by the Earth, as well as the station's travelling through the South Atlantic Anomaly (SAA), splitting the observation into numerous discrete segments.
	
	Data reduction of ObsID 1200120131 was completed using {\sc nicerdas}, a collection of NICER-specific tools, and part of HEASARC. Full Level2 calibration and screening was conducted with {\textit {nicerl2}}, which calibrated, checked the time intervals, merged, and cleaned the data. Barycentric correction was carried out using {\sc barycorr}, then the photon events (all between 0.2-12 keV) were binned to the times of the optical light-curve, and Poissonian errors were applied.

	\section{Results}
	\subsection{Lightcurves \& Discrete Correlation Functions} 
	
	%% Figure: Lightcurve 1
	%\begin{figure*}
	%	\includegraphics[width=\textwidth]{images/lc_optx4.png}
	%	\caption{Small section of optical (top) and X-ray (bottom) lightcurves of MAXI J1820+070. From top to bottom, the optical bands are $g_{s}$ (green), $r_{s}$ (red), $i_{s}$ (dark red), $z_{s}$ (black), and $u_{s}$ (blue). For the optical bands, representative error bars are shown to the left, while the X-ray error bars are plotted in grey. Note the rapid flaring of the source in all bands.}
	%	\label{fig:lc_optx}
	%\end{figure*}
	
	%% Figure: Lightcurve 2
	%\begin{figure}
	%	\includegraphics[width=\columnwidth]{images/lc_flare.png}
	%	\caption{Zoomed-in section of optical lightcurves of MAXI J1820+070, showing an intense, sub-second flare with a peak roughly 50ms wide smaller, quicker flares can be seen to the right. From top to bottom, the optical bands are $g_{s}$ (green), $r_{s}$ (red), $i_{s}$ (dark red), $z_{s}$ (black), and $u_{s}$ (blue).}
	%	\label{fig:lc_flare}
	%\end{figure}
	
    % Figure: Lightcurve + 30s DCF
	\begin{figure*}
		\includegraphics[width=\textwidth]{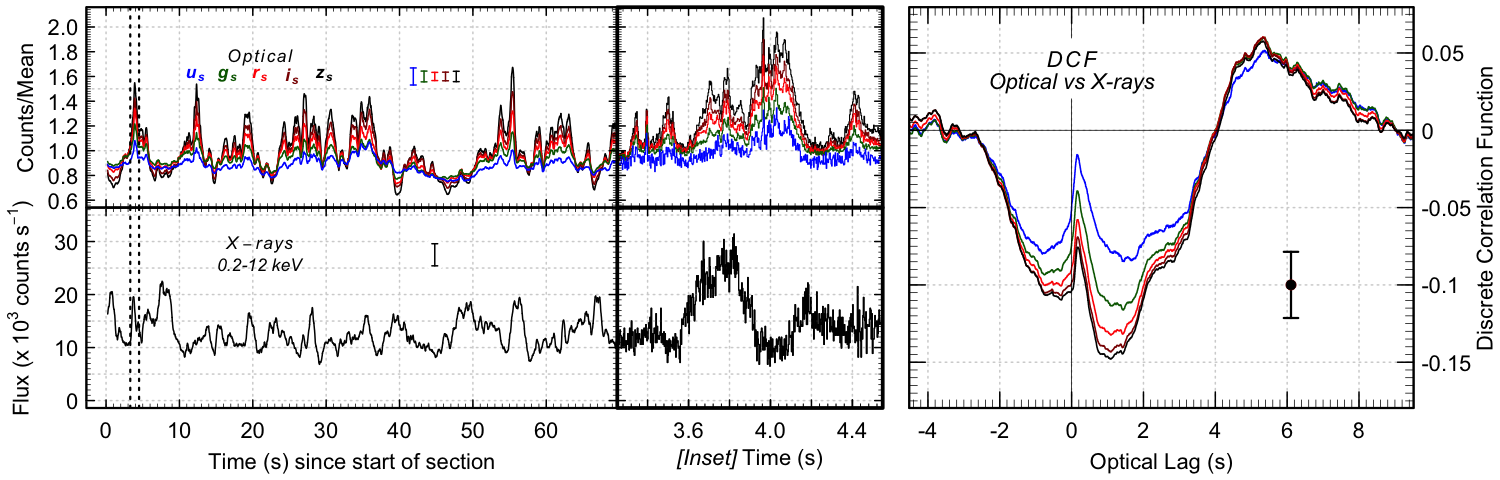}
		\caption{\textbf{Left:} Section of optical (top) and X-ray (bottom) lightcurves. The left panel shows a longer-term trend, binned with a moving average function over 150 points (0.5 seconds), while the right panel shows a zoom-in (illustrated by the dotted line, vertical scales are identical) with no binning. The optical bands are $ugriz_{s}$ (blue/green/red/dark red/black, bottom-to-top at the 4s mark). Representative error bars on individual time bins are shown. Note the rapid red flaring of the source, down to $\sim$\,10ms timescales (e.g. at +3.4\,s and at +3.96\,s). \textbf{Right:} Optical vs X-ray DCF calculated over 52 segments 30s in length. A positive lag here denotes optical emission lagging X-rays. The colours denote which optical band is cross-correlated with X-rays. Around 0 lag, the order of the bands is (bottom-to-top) $z_{s}$--$u_{s}$ in order of wavelength. Error (in all bands) was calculated from bootstrapping over 10,000 iterations.}
		\label{fig:lc_dcf}
	\end{figure*}
	
	J1820 varied rapidly through the course of the night. Sub-second flares were frequent in all bands, of a factor of 2 in optical and 3 in X-ray. Calibrating the flux using HiPERCAM zero points, we found these variations to be up to one magnitude in scale, stronger at longer wavelengths. Some flares were as short as a few bins across ($\sim$10 ms). A representative lightcurve segment can be seen in Figure \ref{fig:lc_dcf}. 
	
	%J1820 varied rapidly through the course of the night. Sub-second flares were frequent in all bands, of a factor of 2 in optical (stronger at longer wavelengths) and 3 in X-ray. Calibrating the flux using HiPERCAM zero points, we found these flares to reach up to 10.5 mag in the $z_{s}$ band. Some flares were as short as a few bins across ($\sim$10 ms). A representative lightcurve segment can be seen in Figure \ref{fig:lc_dcf}. 
	
	%\subsection{Discrete Correlation Functions} \label{dcf}
	The simultaneous nature of the observations also allowed us to create Discrete Correlation Functions (DCFs) measuring the correlation between the optical and X-ray lightcurves as a function of time lag. We split the data into 52 segments, each 30s in duration. After pre-whitening the data to remove any red noise trend \citep{welsh_reliability_1999}, we computed the DCF for each segment \citep{Edelson_DCF_1988}, and the median result was found. Bootstrapping with 10,000 iterations was carried out to find the uncertainties.
	
	The resultant DCFs seen in Figure \ref{fig:lc_dcf} clearly show three main features: an anti-correlation between -3 -- +4 seconds (stronger at longer wavelengths); a positive correlation feature at a lag of $\sim$+165ms in every band; and a hump between +4--9 seconds (positive time lags denote optical lagging X-rays). Each band closely follows the same pattern. Incredibly, analysis of the sub-second peak found it to vary with wavelength; shorter wavelengths peak earlier than longer ones. These shall be discussed in Section \ref{discuss}.

	\subsection{Fourier Analysis} \label{sec:fourier}
	
	% Figure: Coherence Lags
	\begin{figure}
		\includegraphics[width=\columnwidth]{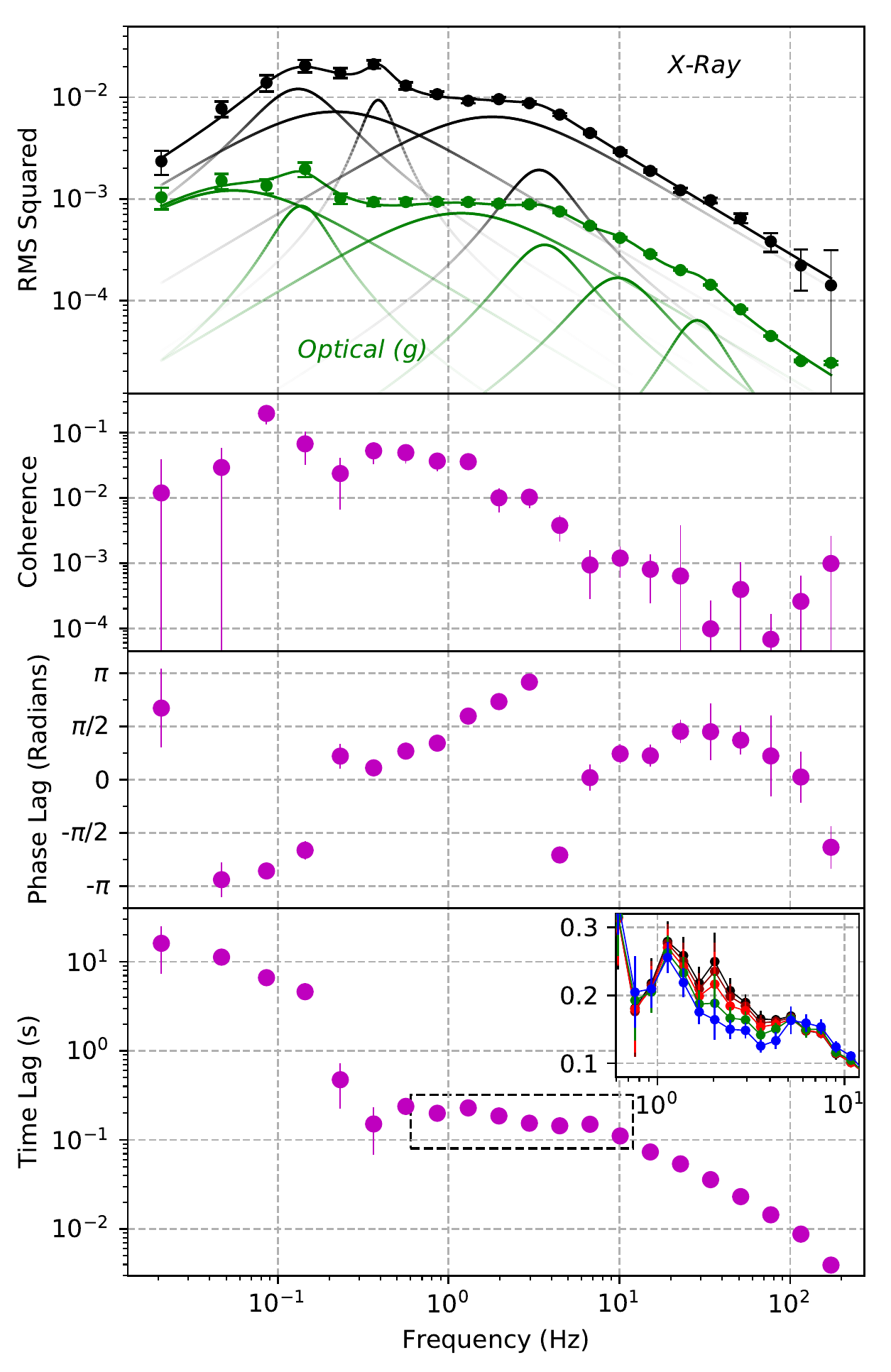}
		\caption{Relations between $g_{s}$ and X-rays: (from top to bottom) X-ray and optical power spectra; intrinsic coherence; phase lags; and time lags. For the last two panels, positive lags mean that $g_{s}$ lags X-rays. We used a logarithmic rebinning factor of 1.5, and data were averaged over segments of 16384 bins (48s). The inset on the last panel shows the time lags with all five bands plotted (over the region shown by the dashed box), with a logarithmic rebinning of 1.2, and colours the same as in Figure \ref{fig:lc_dcf}.}
		\label{fig:coherence}
	\end{figure}
	
	Figure \ref{fig:coherence} shows the power spectra, coherence, and phase and time lags between $g_{s}$ and X-rays. The coherence and phase lags represent the relative magnitude and the phase angle of the complex-valued cross spectrum respectively. This analysis made use of the Stingray\footnote{\url{https://github.com/StingraySoftware/stingray}} python package \citep{Huppenkothen_Stingray_2019}, with errors determined using methods described by \cite{Vaughan_Nowak_1997}. Good Time Intervals (GTIs) were used based on the individual epochs of X-ray observation, then cross spectra were computed over 31 independent lightcurve segments of equal length (16384 bins, 48s) and averaged. RMS squared normalisation was applied to the power spectra \citep{Belloni_Hasinger_Aperiodic_1990}. White noise was fit and removed from the power spectra before calculating the cross spectra. Each optical band showed broadly similar features; the $g_{s}$ band is shown due to its highest signal-to-noise ratio. Our results don't notably change with different X-ray bands.
	
	%Figure \ref{fig:coherence} shows a section of strong coherence between 5 $\times$ 10$^{-2}$ -- 5 Hz, with a particularly high values at around 0.1, and between 0.3 -- 1 Hz, roughly matching lorentzians fitted to the power spectra of the X-ray signal. Beyond this point, the coherence drops, echoing a decrease in the power spectra.
	
	%The phase lags show a continuous upward trend; note that phase lags outside of -$\pi$ and $\pi$ are not physically meaningful, and thus the values can be said to 'wrap around' at these boundaries. The errors are smaller than the point sizes for the majority of the phase lags, but increase dramatically beyond 8 Hz, where the upward trend also stops - again, this takes place just after the X-ray spectrum becomes noise-dominated. A spike in the phase lags can be seen at 0.3 Hz.
	
	For the time lags (\textit{$\tau$ = $\phi$/}2\textit{$\pi$f}, where $\phi$ = phase lag and \textit{f} = frequency), we assume a continuous lag spectrum, and thus allow phase lags outside the range [-$\pi$,+$\pi$]. We first determined that the phase lag around 1 Hz is within [-$\pi$,+$\pi$] (since the time lags shown there are equivalent to those shown in the DCF), and removed the discontinuities by adding 2$\pi$ to the phase lags between 0.03--0.2\,Hz and above 4\,Hz before calculating the time lags - note that this results in only positive time lag values. We also note that the first four points are ambiguous, and could instead be close to -$\pi$ (and thus correspond to negative time lags).
	
	%The time lags are calculated thusly; time lag \textit{$\tau$ = $\phi$/}2\textit{$\pi$f}, where $\phi$ = phase lag and \textit{f} = frequency. We assume a continuous lag spectrum, and thus allow phase lags outside the range [-$\pi$,+$\pi$]: We first determined that the phase lag around 1 Hz is within [-$\pi$,+$\pi$] (since the time lags shown there are equivalent to those shown in the DCF), and removed the discontinuities at 0.03, 0.2 and 4 Hz by adding 2$\pi$ to the required phase lags before calculating the time lags. We note that the first four points are ambiguous, and could instead be close to -$\pi$ (and thus correspond to negative time lags)
	
	%(plotting the data at a higher number of bins was used to confirm this)
	
	%\subsection{Power Spectra and Coherence} \label{discuss:PSCoherence}
	
	The power spectra (PSDs) for both $g_{s}$ and X-ray bands show striking similarities. Lorentzian fitting in both bands found a feature at roughly 0.11 Hz, and similar features between $\approx$\,1--3 Hz. Each of these are associated with significant cross-band coherence. The broad feature seems to dominate the PSDs as they decline at higher frequencies. Note that we see significant optical power up beyond 100 Hz.
	
	There are three significant features of the time lags. The first is between 0.02--0.2\,Hz, where phase lags of close to +$\pi$ indicate that there is some optical component strongly delayed with respect to the X-rays, or, if they are instead -$\pi$, that variations at this frequency are mainly anti-correlated. Following that, there is a significant plateau between 0.5--8\,Hz at $\sim$+165\,ms, corresponding to the peak sub-second lag found in the DCFs. Beyond 8\,Hz, the time lag drops with increasing frequency, consistent with the breaking of the upwards trend in the phase lags.
	
	There are a number of sharp, sudden drops in the coherence; a particular one at 0.2 Hz corresponds to a curious spike in the phase lags and the first discontinuity in the time lags. This frequency, along with frequencies that also feature drops in coherence, tend to coincide with a change in the dominant PSD Lorentzians in both bands. As noted in \cite{Vaughan_Nowak_1997}, this change can cause a loss of coherence; this is especially true if the origins of the Lorentzians are independent \citep{Wilkinson_Uttley_2009}.
	
	%'Wrapping around', as described earlier, would solve this problem, but the large errors and lack of a clear trend present in the phase lags mean that any wrapping would be completely arbitrary.
	
	%With regards to the differences between bands, it is interesting to note that at 0.5 Hz, shorter wavelengths have greater lags; this reverses between 1 -- 3 Hz, where shorter wavelengths have lower lags, is is shown in \ref{fig:dcf_2s_all}.

	\subsection{Wavelength Dependence of sub-second lags}

	% DCF Table
	%\begin{table} 
	%	\centering
	%	\caption{Optical vs X-ray peak and centroid lag for each band.}
	%	\begin{tabular}{cccc}
	%		\hline
	%		Band & $\lambda$ (\AA) & Peak Lag (ms) & Centroid Lag (ms)\\
	%		\hline
	%		$u_{s}$ & 3526 & 154.4 & 169.2\\
	%		$g_{s}$ & 4732 & 164.9 & 172.2\\
	%		$r_{s}$ & 6199 & 173.1 & 178.0\\
	%		$i_{s}$ & 7711 & 170.2 & 186.1\\
	%		$z_{s}$ & 9156 & 176.6 & 186.8\\
	%		
	%		\hline
	%	\end{tabular}
	%	\label{tab:lags}
	%\end{table}
	
			% Figure: Wavelength Peaks
	\begin{figure}
		\includegraphics[width=\columnwidth]{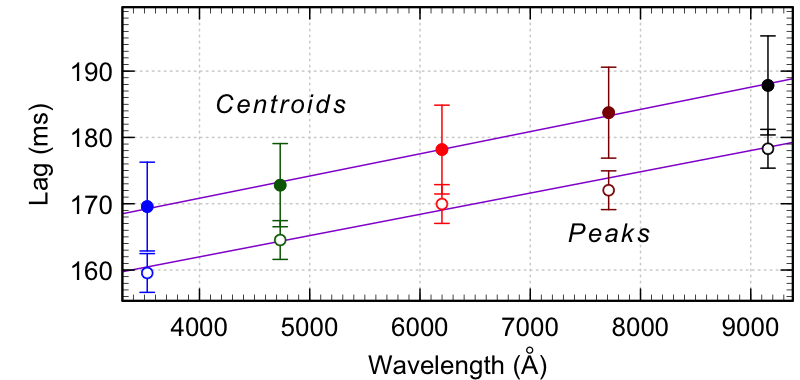}
		\caption{Peak (open circles) and centroid (filled circles) lags of the sub-second correlation peak. Best-fit lines are shown in violet. Colours are as described in Figure \ref{fig:lc_dcf}.}
		\label{fig:wav_peaks}
	\end{figure}

    %% Figure: Coherence Lags
	%\begin{figure}
	%	\includegraphics[width=\columnwidth]{images/Coherence_Zoom.png}
	%	\caption{Zoom-in to the time lags from Figure \ref{fig:coherence}, with all five bands plotted to show the variation by wavelength. Logarithmic rebinning was conducted with a factor of 1.2, and segments of size 12s were used. Colours used are the same as in Figure \ref{fig:lc_dcf}.}
	%	\label{fig:coherence_zoom}
	%\end{figure}
	
    This is the first time that a wavelength dependence has been seen on very short sub-second timescales. To quantify this, we created DCFs using 2s segments of the lightcurve. We then implemented bootstrapping, taking (with replacement) 10,000 samples of segments (with the same sample of segments used for each band). For each iteration, we calculated a mean DCF, and recorded the peak of the sub-second lag ($\pm$half a lag bin, i.e. $\sim$3 ms).
	
	A linear trend was then fitted between peak lag and wavelength for each iteration, and the mean, 16\%, and 84\% values were calculated to be $3.15^{+1.43}_{-1.57}$ $\mu$s \AA$^{-1}$. Centroids were calculated using methods similar to \citet{koratkar_structure_1991} and \citet{gandhi_elevation_2017}; for each iteration, we summed over all lags where the DCF coefficients were 80\% of peak value. The standard deviation was calculated for the entire distribution. A linear fit gave a slope of 3.35 $\pm$ 3.03 $\mu$s \AA$^{-1}$. The results are shown in Figure \ref{fig:wav_peaks}.
	
    When we plot time lags for each optical band, this same qualitative wavelength dependence is present between Fourier frequencies of 1--5 Hz. This can be seen in Figure \ref{fig:coherence}.
	
	%To do this, we measured both the peaks and centroids of the sub-second feature seen in Figure \ref{fig:lc_dcf}. For the peaks, we created DCFs using 2s long segments of the lightcurve;, averaged them, interpolated between points to smooth the line, took a ten-point moving average function, and found the highest point of the resultant curve ($\pm$half an original lag bin size, i.e. $\sim$3 ms). Centroids were calculated using methods from \citet{koratkar_structure_1991} and \citet{gandhi_elevation_2017}; we summed over all lags where the DCF coefficients were 80\% of peak value, uncertainties were computed by simulating 10$^3$ DCFs, using the standard error of the original coefficients and assuming a normal distribution, and then the standard deviation was calculated. A linear fit gives a slope of 3.43 $\pm$ 1.27 $\mu$s \AA$^{-1}$ for the peaks, and 3.47 $\pm$ 0.24 $\mu$s \AA$^{-1}$ for the centroids (Figure \ref{fig:wav_peaks}). Fourier analysis also shows this lag; plotting time lags for each optical band individually reveals the wavelength dependence (Figure \ref{fig:coherence}).

	% Alternative Discussion
	
	\section{Discussion} \label{discuss}

	%\subsection{Observed Features and Previous Observations}
	
	Our observations have highlighted several intriguing features that any interpretation needs to explain. These include: The sub-second lag; the wavelength dependence of this lag; the broad anti-correlation; the slow $\sim$\,+5\,s positive correlation; the phase lags; and the red flares seen in the lightcurve.
	
	Many of these features have been seen before in LMXBs; sub-second correlations have previously been found in XTE J1118+480  \citep{kanbach_correlated_2001}, GX 339-4 \citep{Gandhi_Correlations_2008, Vincentelli_GX339_2018}, V404 Cyg \citep{gandhi_elevation_2017}, and Swift J1357.2-0933 \citep{Paice_J1357_2019}. XTE J1118+480 also showed the wavelength-dependence of such a feature, albeit on longer timescales  \citep{hynes_remarkable_2003} and between the UV and X-ray. Meanwhile, broad anti-correlations are common in LMXBs \citep{kanbach_correlated_2001, durant_swift_2008, Pahari_BWCir_2017} as well as red flares \citep{Gandhi_Correlations_2008, gandhi_furiously_2016}. %Fourier analysis has also been carried out on GX 339-4 by \citet{gandhi_rapid_2010} and \citet{Malzac_GX339_2018}, finding similar results those those shown in Figure \ref{fig:coherence}.
	The very rapid times associated with these features allow direct optical probes of processes very close in to the central engines in LMXBs. The high time resolution and wavelength coverage offered by HiPERCAM/GTC, together with the X-ray bright-source throughput of NICER, is unprecedented, and allows us to investigate the models suggested for these earlier sources at a greater resolution than before.

	%\subsection{Possible Models; Jet Emission, Doppler Boosting and Hot Accretion Flow}
	
	%The multiple features we see here require several different models to explain them. In this section, we put forward phenomena that could be present in the system, and discuss how they might give rise to the observed features.
	
	A scenario that explains many of the observed features is synchrotron emission from internal shocks within a relativistic compact jet. In this model, infalling matter emits X-rays close to the black hole, and plasma shells are accelerated along a jet. These shells, with speeds dependent on the variable inflow of matter, would then collide and shocked material would emit in the optical and infrared \citep{Malzac_Shocks_2013}. The +165\,ms lag would thus be the average travel time for the material between the X-ray corona (analogous to the jet base, \citealt{Markoff_Coronae_2005}) and the optical emitting regions of the jet; assuming material travelling at light speed, this corresponds to a maximum distance of roughly 4650 R$_{G}$ \textit{($\equiv$ GM/c$^{2}$)} for a $\sim$\,7 M$_{\sun}$ black hole \citep{Torres_Dynamical_2019}. An optical lag of $\approx$\,0.1\,light-seconds appears, in fact, to be common in LMXBs in the hard state (modulo factors of $\mathcal{O}(1)$ related to plasma velocity and viewing geometry), and is likely to be constraining the elevation of the first plasma acceleration zone above the black hole \citep{gandhi_elevation_2017}. 
	
	What can the wavelength dependence tell us? A linear dependence of characteristic emission wavelength with distance from the central compact object (hence, time lag) is far from a novel result, and is in fact a key prediction for the optically-thick emitting zones in compact jet theory \citep{blandford_relativistic_1979}; however, our data only show a roughly 12\,\% change in lag over the probed optical wavelength range; this is too shallow to be explained by such a linear dependence. Similarly, our slope of $\sim$3.25 $\mu$s \AA$^{-1}$ is a factor of 50 smaller from that found in XTE J1118+480 by \citet[][160 $\mu$s \AA$^{-1}$]{hynes_remarkable_2003}. This is too great a difference to be due to simple length (and thus mass) scaling, and is consistent with the idea that there are other factors that affect these lags, such as inner accretion disc radius and magnetic field strength (\citealt{Russell_Evolving_2013}). %Furthermore, our observations likely probed optically-thin emission near the outburst peak \citep{}{RUSSELL ATEL}. 
	
	%It is plausible that this first plasma acceleration zone is itself somewhat elongated. 
	
	Instead, we may be seeing the first signs of stratification within the innermost jet emitting zones. Emission here is expected to be optically-thin \citep[e.g.][]{markoff_jet_2001, Russell_1820_ATel_2018}, but this is likely only true {\em on average}; colliding shocks would create a distribution of velocity shears \citep{Malzac_Shocks_2013}, with faster shocks peaking at higher spectral frequencies (due to self-absorption) and at slightly shorter lags than slower shocks -- in qualitative agreement with the wavelength dependent trends shown in Figs. \ref{fig:coherence} \& \ref{fig:wav_peaks}. This `first shock dissipation zone' has been modelled before (\citealt{Ceccobello_RMHD_2018} and references), but the precise time-resolved dissection of data that we present is new, and further specific modelling of the physics behind these lags is needed.
	
	%\textbf{Further to this, we may also be seeing the first signs of jet stratification; as the jet propagates, it becomes optically thin to progressively lower energies. Non-thermal emission of shorter wavelengths would thus occur closer to the jet base, also creating a dependence we qualitatively see here.}

    %could be explained by the difference in velocity between plasma shells; we may be starting to see stratification of the first shock dissipation zones, which are only expected to be optically thin on average. The observed lag trend of small increases with wavelength agrees qualitatively with a scenario in which shorter lags arise from larger velocity shear amongst the colliding shocks. These shocks result in marginally faster dissipation than collisions arising from lower velocity shear, which therefore lag further. The observed time differences with wavelength may potentially allow the first direct constraints on the velocity shear in this zone.
	
	Our low-frequency phase lags support multiple models. The phase lags encompass a range of absolute values between $\sim$\,$\pi$/2 and $\pi$, and are likely to comprise a mix of components. The magnitudes of the corresponding time lags of $\sim$\,few--10\,s are associated with both the anti-correlation seen in the DCFs as well as the slower positive correlation at $\sim$\,+5\,s. A phase lag magnitude of $\pi$ %with respect to X-rays (recall that for phase lags, +$\pi$ $\equiv$ -$\pi$), they could 
	corresponds to the observed anti-correlation. This could arise from Doppler Boosting within the jet; for a given inclination, as the jet Lorentz factor increases, the apparent luminosity of a jet \textit{decreases} due to relativistic beaming (see \citealt{Malzac_GX339_2018}). This leads to apparently less jet optical and infrared flux along the line of sight. Both a high inclination angle or a high jet Lorentz factor could play roles here. Alternatively, a hot flow scenario could also provide a self-consistent explanation  \citep{Narayan_Yi_ADAF_1994}; this is suggested to be present in this source by both \citet{Veledina_Polarisation_2019} and \citet{Kajava_Dips_2019}. Here, an increase in mass accretion rate would lead to increased X-ray flux, and a higher level of synchrotron self-absorption. The latter would then lead to a drop in the optical emission \citep{veledina_accretion_2013}.
	%\citet{Veledina_Precession_2013} also noted that at higher inclinations (implied by the doppler boosting scenario), a precessing hot flow would create a QPO in both the optical and X-ray power spectra, which is a possible feature that we see in Figure \ref{fig:coherence}; whether this feature is significant, and the strength of the consistency with this model, would be a topic for future work.
	%If our low-frequency time lags, then they are...
	%If our low-frequency phase lags are +$\pi$, then they are consistent with is a
	Finally, the long positive correlation on optical lags of $\sim$\,+5--15\,s could originate in disc reprocessing, suggested by both \citet{Paice_J1820_2018} and \citet{Kajava_Dips_2019}. Multiwavelength modeling and assessment of this scenario will help to constrain the disc extension, and should be carried out in future work. %The time lags of 5 -- 15s that they would give are similar to the lag of the extended feature seen in Figure \ref{fig:lc_dcf}.
	
	In Sec. \ref{sec:Intro}, we noted that LMXB emission is considered to be a mixture of processes. The data is not only consistent with elements of each of those, but implies multiple components; from the Fourier analysis, features above 0.2 Hz would be caused by a jet, while those below would be related to accretion variability from the hot flow and disc (\citealt{Wilkinson_Uttley_2009}, \citealt{Churazov_Soft_2001}, \citealt{done_modelling_2007}).

	%Finally, it should be noted that a multi-component interpretation is also consistent with the Fourier components mentioned in Section \ref{sec:fourier}. Features above 0.2 Hz would be caused by jet emission, while those below would be related to accretion variability leading to the hot flow (\citealt{Wilkinson_Uttley_2009}, \citealt{Churazov_Soft_2001}, \citealt{done_modelling_2007}).

\vspace*{0.3cm}
\noindent
%section{Conclusions}
    MAXI\,J1820+070 was the brightest LMXB transient in 2018, and studies of its multiwavelength emission will undoubtedly continue to prove valuable. Here, we have presented a first look in the richness of information available on millisecond timescales. We find a novel multiband time-lag trend with wavelength, but also noted that many results %here, particularly in regards to the sub-second lag and the Fourier components, 
    echo similar findings in systems like GX 339-4 and V404 Cyg. Indeed, it increasingly seems that time and length scales are similar across LMXBs. Testing this trend through analysis of future LMXB sources should prove most interesting; tests that, with this newest generation of telescopes, we now have the ability to carry out better than ever before.

	\section*{Acknowledgements}

	We acknowledge support from STFC and a UGC-UKIERI Thematic Partnership. JAP is part supported by a University of Southampton Central VC Scholarship, and thanks D Ashton for spectral timing help. Observations were made with the GTC telescope (Spanish Observatorio del Roque de los Muchachos, Instituto de Astrof\'\i{}sica de Canarias), under Director's Discretionary Time. JM acknowleges financial support from PNHE in France, OCEVU Labex (ANR-11-LABX-0060), and the A*MIDEX project (ANR-11-IDEX-0001-02) funded by the ``Investissements d’Avenir” French government program managed by the ANR. TS thanks the Spanish Ministry of Economy and Competitiveness (MINECO; grant AYA2017-83216). AV acknowledges the Academy of Finland grant 309308. HiPERCAM and VSD funded by the European Research Council (FP/2007--2013) under ERC-2013-ADG grant agreement no. 340040. SMARTNet helped to coordinate observations. We also thank the referee for their valuable comments. We have made use of %NASA's Astrophysics Data System, and data, 
	software and web tools from the High Energy Astrophysics Science Archive Research Center (HEASARC).%, a service of the Astrophysics Science Division at NASA/GSFC and of the Smithsonian Astrophysical Observatory's High Energy Astrophysics Division.

	%%%%%%%%%%%%%%%%%%%%%%%%%%%%%%%%%%%%%%%%%%%%%%%%%%
	
	

	\bibliography{ref}
	\bibliographystyle{mnras}

	% \begin{thebibliography}{99}
	% \bibitem[\protect\citeauthoryear{Krimm et al.}{2011}]{Krimm et al. 2011}
	% Krimm H. A. et al., 2011, The Astronomer's Telegram, 3138, 1 
	% \bibitem[\protect\citeauthoryear{Others}{2013}]{Others2013}
	% Others S., 2012, Journal of Interesting Stuff, 17, 198
	% \end{thebibliography}
	
	%%%%%%%%%%%%%%%%%%%%%%%%%%%%%%%%%%%%%%%%%%%%%%%%%%
	
	%%%%%%%%%%%%%%%%% APPENDICES %%%%%%%%%%%%%%%%%%%%%
	
	%\appendix
	
	
	% If you want to present additional material which would interrupt the flow of the main paper, it can be placed in an Appendix which appears after the list of references.
	
	%%%%%%%%%%%%%%%%%%%%%%%%%%%%%%%%%%%%%%%%%%%%%%%%%%

	% Don't change these lines
	\bsp % typesetting comment
	\label{lastpage}
\end{document}